Paul Schmitt,[1,2] Pallabi Paul,[1,2] Weiwei Li,[3,4] Zilong Wang,[3,4] Christin David,[5] Navid Daryakar,[5] Kevin Hanemann,[1] Nadja Felde,[1] Anne-Sophie Munser,[1,2] Matthias F. Kling,[3,4,6,7] Sven Schröder,[1] Andreas Tünnermann,[1,2] and Adriana Szeghalmi[1,2,*]

# Linear and Nonlinear Optical Properties of Iridium Nanoparticles by Atomic Layer deposition

[1] Fraunhofer Institute for Applied Optics and Precision Engineering IOF, Center of Excellence in Photonics, Albert-Einstein-Straße 7, D-07745 Jena, Germany

[2] Institute of Applied Physics and Abbe Center of Photonics, Friedrich Schiller University Jena, Albert-Einstein-Straße 15, D-07745 Jena, Germany

[3] Physics Department, Ludwig-Maximilians-Universität Munich, Am Coulombwall 1, D-85748 Garching, Germany

[4] Max Planck Institute of Quantum Optics, Hans-Kopfermann-Straße 1, D-85748 Garching, Germany

[5] Institute of Condensed Matter Theory and Optics and Abbe Center of Photonics, Friedrich Schiller University Jena, Max-Wien Platz 1, D-07743 Jena, Germany

[6] SLAC National Accelerator Laboratory, 2575 Sand Hill Rd, Menlo Park, CA 94025, USA

[7] Department of Applied Physics, Stanford University, 348 Via Pueblo, Stanford, CA 94305, USA

**Abstract:** Nonlinear optical phenomena enable novel photonic and optoelectronic applications. Especially metallic nanoparticles and thin films with nonlinear optical properties offer the potential for micro-optical system integration. For this purpose, new nonlinear materials need to be continuously identified, investigated, and utilized for nonlinear optical applications. While noble metal nanoparticles, nanostructures, and thin films of Ag and Au were widely studied, iridium (Ir) nanoparticles and ultra-thin films have not been investigated yet. Here, we present a combined theoretical and experimental study on the linear and nonlinear optical properties of Ir nanoparticles deposited by atomic layer deposition (ALD). Linear optical constants, i.e., the effective refractive index n and extinction coefficient k, were evaluated at different growth stages of nanoparticle formation. Both linear and nonlinear optical properties of these Ir ALD coatings were calculated theoretically using Bruggeman and Maxwell-Garnett theories. The third-order susceptibility of Ir nanoparticle samples was experimentally investigated using the Z-scan technique. Exemplary, for an Ir ALD coating with 45 cycles resulting in small Ir nanoparticles, the experimentally determined nonlinear third-order susceptibility is about $\chi_{Ir}^{(3)} = (2.4 - i2.1) \times 10^{-17}$ m²/V² at the fundamental wavelength of 700 nm. By using the nonlinear Maxwell-Garnett theory with Miller's rule, this nonlinear susceptibility is calculated $(3.55 + i15.70) \times 10^{-11}$ m²/V² and for nonlinear Bruggeman theory it results in $(1.06 + i2.08) \times 10^{-14}$ m²/V² at 233 nm. The strong increase is due to the proximity to the Mie resonance of Ir nanoparticles.

## 1 Introduction

Materials with strong nonlinear optical properties (NOP) enable novel photonic and optoelectronic applications for optical communication and data processing [1,2], optical data storage [3], imaging [4,5], optical limiters [6], ultrafast switches [7], photocatalysis [8], etc. While bulk crystals [9], such as β-BaB$_2$O$_4$ and LiNbO$_3$, have been widely exploited for different nonlinear processes, nanomaterials, for instance, metallic nanoparticles (NP) or nanostructures [10] and two-dimensional (2D) materials [11] could exhibit significant NOP due to the plasmonic enhancement and quantum size effects, enabling the development of complex nanostructured systems for micro-optical system integration. Among them, localized nanoscopic light sources in fluorescence spectroscopy from gold (Au) nanoring structures incorporated with LiNbO$_3$ [12], improved third-order nonlinearities of Au NP in sapphire matrix [13], quantum size effect-based enhancement of third-order nonlinearities in silver (Ag) NPs embedded in silica glass [14], or promising surface second-harmonic generation from Au, Ag, Al, W metal stacks [15] are recent developments. Other exciting materials for nonlinear applications include oxide thin films and multilayers, e.g., ITO [16], $Al_2O_3$/ZnO [17], and $Al_2O_3$/$TiO_2$/$HfO_2$ [18] based optical metamaterials.

While significant interests for nonlinear processes in metallic nanomaterials are primarily in gold or silver with their plasmonic resonances residing in the visible, other metals are less studied. Depending on specific applications, for instance, superior optical performance in the UV spectral regime, further enhancement of NOP and new nonlinear materials, e.g., various metallic NPs, embedded NPs, and nanostructured metasurfaces, need to be identified and investigated. Iridium is an exceptionally stable metallic material with high thermal stability and oxidation resistance. In optics, metallic Ir thin films are usually applied as X-ray mirrors [19,20] at grazing incidence due to their high density [21]. Further applications include Fresnel zone plates [22], metal wire grid polarizers in the UV [23], and protection layers [24]. Recently, we demonstrated the application potential of stable and highly reflective Ir mirror coatings for the infrared (IR) spectral range [25]. Iridium-complexes with suitable ligands show second-order NOP and two-photon absorption properties [26]. The linear optical properties of thick metallic Ir films have been reported [25,27,28]; however, ultrathin and partially transparent NP assemblies mostly relevant for nonlinear optical applications have not been investigated yet.

This article discusses the linear and nonlinear optical properties of Ir nanoparticles. The samples were prepared by atomic layer deposition (ALD), based on sequential, self-limiting surface reactions, allowing for a precise sample thickness control [29,30]. Iridium nano-sized islands are formed prior to the growth of a continuous thin film, and by controlling the reaction cycles, NPs of different sizes and surface coverage can be yielded [31]. Consequently, the linear and NOP of Ir depend strongly on the stage of growth. Here, a combination of spectroscopic and microscopic tools, including spectrophotometry (SP), spectroscopic ellipsometry (SE), and angle-resolved scattering (ARS), was applied, complemented by X-ray reflectometry (XRR), white light interferometry (WLI), and scanning electron microscopy (SEM) to determine the structural and optical properties of Ir NPs. The NOP of these Ir ALD coatings are calculated using Maxwell-Garnett theory for their third-order susceptibility [32] and compared with experimentally obtained values using the femtosecond Z-scan technique.

## 2 Methods

**Deposition**. Amorphous fused silica (FS) with an ultra-flat surface was used as substrates. The typical AFM (1 × 1 µm²) root-mean-square (rms) surface roughness is about 0.26 nm. Their cleaning was performed with a multi-stage, ultrasonic-assisted bath cleaning system (Elma Schmidbauer, Singen, Germany) with alternating surfactants and water ($H_2O$) baths, concluded by a deionized, ultra-pure $H_2O$ bath.

The depositions were performed with a commercial SunALE R-200 Advanced ALD system (Picosun Oy, Masala, Finland) with iridium(III) acetylacetonate (Ir(acac)$_3$) and molecular oxygen ($O_2$) as precursors. A heatable wafer chuck ensures a substrate temperature of 380 °C. One Ir ALD cycle consists of 6 s of Ir(acac)$_3$ pulse, 60 s of purge, 2 s of $O_2$ pulse, and 6 s of purge with molecular nitrogen ($N_2$) as purging gas. By tuning the growth cycles (30-400 cycles), Ir coatings with NPs and ultrathin films are grown with an effective thickness varying between 1.6 nm and 25 nm. Details of this thermal ALD process and the formation of poly-crystalline Ir thin films can be seen in previously published articles [25,31,33].

**Characterization**. Structural properties of the Ir coatings were determined by X-ray reflectometry (XRR), white light interferometry (WLI), and scanning electron microscopy (SEM). For the XRR measurements, a D8 Discover diffractometer (Bruker AXS, Karlsruhe, Germany) with Cu Kα radiation (λ = 0.154 nm), 40 kV cathode current, and 40 mA acceleration voltage in Bragg-Brentano geometry was used. A NewView 7300 system (Zygo, Middlefield, CT, USA) with 50x magnification was used for WLI measurements. SEM images were obtained with a field-emission SEM Hitachi S-4800 (Hitachi, Tokyo, Japan) using 0.7 kV acceleration voltage and 2.0 – 3.2 mm working distance. The open-source image processing program ImageJ [34] was used to analyze the SEM images.

The linear optical properties were determined using spectrophotometry (SP), variable-angle spectroscopic ellipsometry (SE), and angle-resolved scattering (ARS). Reflectance R, transmittance T, and optical losses OL = 1-R-T from 200 nm to 2200 nm were measured with a two-beam spectrophotometer Lambda 900 (PerkinElmer, Waltham, MA, USA) at 6° angle of incidence. The ellipsometric parameters Ψ and Δ from 190 nm to 980 nm were determined with a spectroscopic ellipsometer SE850 DUV (Sentech Instruments, Berlin, Germany) at angles of incidence between 40° and 70°. The optical constants of Ir, namely the effective refractive index n and extinction coefficient k, were evaluated using the SpetraRay/4 (Sentech Instruments, Berlin, Germany) software package by fitting the Ψ and Δ data. Therefore, a three-layer model consisting of an FS substrate, an effective Ir layer using a Drude-Lorentz model with five oscillators, and the surface roughness as an effective-medium-approximation (EMA) top layer according to Bruggeman were applied. As determined by XRR, the thicknesses of the effective Ir layer and EMA layer were kept constant during the fit procedure. ARS was performed using an ALBATROSS-TT system (Fraunhofer IOF, Jena, Germany) at normal incidence and 405 nm wavelength with a 3 mm illumination spot in forward and backward scattering [35].

Nonlinear optical properties were investigated using the Z-scan technique. A home-built non-colinear optical parametric amplifier (NOPA) driven by a Ytterbium-based fiber laser (Active Fiber Systems, Jena, Germany) with a 100 kHz repetition rate was used as the light source. Then, we selected the fundamental wavelength at 700 nm using a 40 nm bandpass filter resulting from a typical 100 fs pulse. An achromatic lens focused the laser light with a focal length of 10 mm, and the beam waist at the focus position was about 48 µm. The open aperture (OA) and closed aperture (CA) Z-scan measurements of the Ir NP layers were performed simultaneously under an average incident power of 80 mW, corresponding to a laser intensity of about 218 GW/cm$^2$ at the focus position. Here, the laser intensity is chosen so that the signals from the Ir NP samples can be clearly distinguished from substrate contributions. The samples were mounted on a motorized linear translation stage, which allowed scanning through the focus of the beam. In order to reduce the thermal effects accumulated on the sample surface, a mechanical shutter was placed in front of the sample. This shutter was switched on for 100 ms for each Z-position and then switched off while the sample was moved to the next Z-position. The OA and CA signals were detected using two biased Si photodiodes (DET10A2, Thorlabs, Newton, NJ, USA) and then demodulated with two lock-in amplifiers (SR830, Stanford Research Systems, Sunnyvale, CA, USA) at laser repetition rate.

**Simulation**. The Maxwell-Garnett (MG) and Bruggeman (BG) theories [32,36] were employed to model the optical properties of the Ir coatings, using their layer thickness and surface coverage as determined by XRR and SEM, respectively. Standard MG assumes spherical material inclusions (Ir) inside a host matrix (air) with a volume fill fraction f in quasi-static and dipolar approximation. However, particle-particle interactions are not considered, so that MG loses its validity with increasing fill fraction. The local field enhancement factor η for spherical inclusions is

$$\eta = \frac{\varepsilon_1 - \varepsilon_2}{\varepsilon_1 + 2\varepsilon_2} \qquad 1$$

with the permittivity of the spherical Ir inclusions $\varepsilon_1$ and air as host material $\varepsilon_2$. The effective permittivity $\varepsilon_{MG}$ in the MG theory is calculated from

$$\varepsilon_{MG} = \varepsilon_2 \frac{1+2f\eta}{1-2f\eta} \qquad 2$$

and simulates the permittivity for a homogeneous material consisting of the two materials. This strict quasi-static approximation can be lifted by considering a dipolar Mie term [37], which is relevant for larger particle sizes.

On the other hand, the BG theory allows more complex mixtures of the constituents by not assuming a specific shape but including a polarization factor p. Moreover, there is no distinction between the composite into a host and inclusion material. The effective permittivity $\varepsilon_{BG}$ is obtained from

$$\sum_{m=1}^{2} \frac{\varepsilon_m - \varepsilon_{BG}}{\varepsilon_{BG} + p(\varepsilon_m - \varepsilon_{BG})} = 0 \qquad 3$$

with an analytical solution for two constituent materials. The corresponding complex refractive index ñ = n+ik is obtained from $\tilde{n} = \sqrt{\varepsilon_{EMA}}$. In thin-film calculations based on Fresnel coefficients, this will be used to include the layer thickness of different Ir ALD coatings.

Nonlinear extensions for both MG [38,39] and BG [40,41] theories exist and were adapted to analyze the ultrathin Ir NPs layers on FS in air. First, the Mie resonances of Ir NPs with frequencies $\omega_0(R)$ depending on the particle size d = 2R were calculated. Second, we adjusted the third-order electron response to a third-order nonlinear bulk susceptibility in Lorentzian form [42]

$$\chi_1^{(3)}(R) = \frac{a}{D(\omega,R)^2 D(\omega,R) D(-\omega,R)} = \frac{a}{D(\omega,R)^2 |D(\omega,R)|^2} \qquad 4$$

with the denominator $D(\omega, R) = \omega_0^2 - \omega^2 - i\Gamma\omega$ [39]. The contributions of air as a host material and FS as substrate with $\chi_2^{(3)} = 1.77 \times 10^{-25}$ m²/V² and $\chi_{FS}^{(3)} = 2.5 \times 10^{-22}$ m²/V², respectively, are small compared to the $\chi_1^{(3)}(\omega \to 0)$ limit. This is consistent with experimental observations comparing measurements in vacuum and air. The low-frequency limit known as Miller's rule [43] allows estimating the nonlinear coefficient a in varying host materials, which leads to $\chi_1^{(3)}(\omega \to 0) = 1.98 \times 10^{-19}$ m²/V² for a particle size of 5 nm. In the composite material, assuming nonlinear properties from both Ir NPs and the host environment, the nonlinear MG expression becomes [38,39]

$$\chi_{\text{eff}}^{(3)} = f\nu^2|\nu|^2\chi_1^{(3)} + (1 - f + xf)\varrho^2|\varrho|^2\chi_2^{(3)} \qquad 5$$

$$\text{with } \nu = \frac{\varepsilon_{\text{MG}} + 2\varepsilon_2}{\varepsilon_1 + 2\varepsilon_2}, \varrho = \frac{\varepsilon_{\text{MG}} + 2\varepsilon_2}{3\varepsilon_2}, \text{ and} \qquad 6,7$$

$$x = \frac{8}{5}\eta^2|\eta|^2 + \frac{6}{5}\eta|\eta|^2 + \frac{2}{5}\eta^3 + \frac{18}{5}(\eta^2 + |\eta|^2) \qquad 8$$

By having third order susceptibility of bulk and Ir coatings the nonlinear refractive index $n_2$ and absorption coefficient β has been calculated in the table 2, according to the formalism established previously [44]. All these factors describe different enhancement processes of the system at different orders. Finally, we compute the overall optical properties from $\vec{P} = \varepsilon_0\varepsilon_{\text{tot}}(\omega, R)\vec{E}(\omega)$ with

$$\varepsilon_{\text{tot}}(\omega, R) = \varepsilon_{\text{MG}}(\omega) + 3\chi_{\text{eff}}^{(3)}(\omega, R)|\vec{E}|^2 + \chi_{\text{eff}}^{(3)}(3\omega, R)\vec{E}\vec{E} \qquad 9$$

including Kerr nonlinearities at the fundamental wavelength (FH). The third harmonic term is considered separately for evaluating third-harmonic generation (THG) processes as it gives vanishing contributions far off resonance. This material model and the scattering matrix approach are used to describe the system of effectively homogeneous layers [32].

## 3  Results and Discussions

**Linear optical properties**. Figure 1 shows the measured reflectance R, transmittance T, and calculated optical losses OL=1-R-T spectra of Ir ALD coatings on FS substrates, with ALD cycles varying from 30 to 60 cycles. With increasing ALD growth cycles, initial changes in the optical properties emerge in the deep ultraviolet (DUV) spectral range for small cycle numbers. While the reflectance of the Ir coatings up to 45 cycles (light green) is nearly identical to the bare FS substrate (red), the transmittance around 200 nm wavelength decreases, and consequently, the optical losses increase. These coatings are in the regime of nanoparticle growth, as confirmed by SEM investigations [31]. With several tens of cycles, the grown Ir NPs have a weak Mie scattering of localized surface plasmons, and more intense interband transitions dominate in the DUV spectral range. About 60 Ir ALD cycles lead to about 20% optical losses around 230 nm wavelength with very low scattering losses (see Table1).

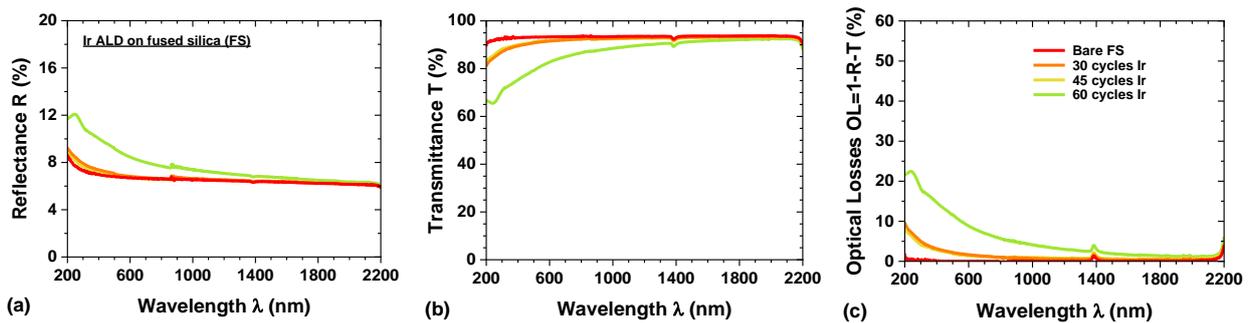

**Figure 1**. (a) Reflectance, (b) transmittance, and (c) optical losses of iridium coatings deposited on fused silica (FS) using 30 - 60 atomic layer deposition (ALD) cycles.

From about 100 Ir ALD cycles, the reflectance increases significantly in the whole spectral range from 200 – 2200 nm (see SI). As an indicator for the formation of continuous metallic films, such an increase in reflectance was also reported in evaporated Au [45] and Ag [46] coatings. Iridium coatings with more than 75 cycles exhibit pronounced OL of about 10 – 40 % in the IR spectral range (see SI). The initial Ir ALD thin-film formation follows the Volmer-Weber growth type. Similar to physical vapor deposition (PVD) processes, the small Ir NPs continuously grow, coalesce, form a coherent network, and finally form a closed ultrathin film. The size of the isolated NPs ranges from 2 – 20 nm [31], which results in absorption and low Mie scattering in the UV spectral range. The total scattering TS of these Ir coatings, depending on their number of ALD cycles, is provided in the SI. This total scattering is more pronounced at 100 ALD cycles when an interconnected network has formed. However, the total scattering with TS < 0.1 % is negligible compared to the overall optical losses. Therefore, the exceptionally high OL in the 100 cycles sample could be mainly due to the light confinement effect, particularly in the IR region. For comparison, the OL and TS of the Ir ALD coatings, as well as their effective layer thickness, density, and surface roughness, are listed in Table 1.

**Table 1**. Properties of Ir ALD coatings deposited on FS. The effective Ir layer thickness and density were analyzed using X-ray reflectometry (XRR); white light interferometry (WLI) and XRR for the surface roughness.

| Sample ID | Number of ALD Cycles | Ir Thickness XRR (nm) | Ir Density XRR (g/cm³) | Ir Surface Roughness (nm) | | Optical Losses @ 405 nm (%) | Total Scattering @ 405 nm (ppm) |
|---|---|---|---|---|---|---|---|
| | | | | XRR | WLI | | |
| 1 | 30 | 1.6 ± 1.0 | 8.5 ± 2.0 | 0.4 ± 0.2 | 0.5 ± 0.1 | 3.1 ± 0.5 | 174 ± 17 |
| 2 | 45 | 2.0 ± 1.0 | 4.1 ± 2.0 | 0.8 ± 0.2 | - | 2.7 ± 0.5 | 40 ± 4 |
| 3 | 60 | 3.0 ± 1.0 | 11.8 ± 1.0 | 1.0 ± 0.2 | 0.5 ± 0.1 | 14.5 ± 0.5 | 252 ± 25 |
| 4 | 75 | 4.0 ± 1.0 | 18.0 ± 1.0 | 1.2 ± 0.2 | - | 24.2 ± 0.5 | - |
| 5 | 100 | 5.7 ± 1.0 | 20.1 ± 0.5 | 0.9 ± 0.2 | 0.6 ± 0.1 | 37.5 ± 0.5 | 711 ± 71 |
| 6 | 150 | 8.7 ± 1.0 | 22.5 ± 0.5 | 1.4 ± 0.2 | 0.9 ± 0.1 | 37.8 ± 0.5 | 79 ± 8 |
| 7 | 200 | 11.7 ± 1.0 | 22.3 ± 0.2 | 1.2 ± 0.2 | 0.9 ± 0.1 | 36 0 ± 0.3 | 236 ± 24 |
| 8 | 250 | 14.3 ± 1.0 | 22.3 ± 0.2 | 1.1 ± 0.2 | 0.4 ± 0.1 | 34.8 ± 0.3 | 217 ± 22 |
| 9 | 400 | 24.7 ± 1.0 | 22.4 ± 0.1 | 0.9 ± 0.2 | 0.4 ± 0.1 | 31.5 ± 0.3 | 87 ± 9 |

The effective linear refractive index n and extinction coefficient k of the Ir coatings were determined by spectroscopic ellipsometry. The accuracy of this ellipsometry analysis is supported by variable angle spectroscopic ellipsometric measurements as summarized in the SI materials. A good agreement between the measured and fitted Ψ and Δ parameters can be seen at angles of incidence between 40° and 70° (see SI). For thin and fully closed Ir films, the optical constants reach nearly the bulk values. For better visualization, Fig. 2.a and c demonstrate the dispersion profiles of Ir coatings up to 60 ALD cycles. While the system is in the NP growth regime, the optical constants are relatively low. As the Ir NP system approaches the percolation threshold, a significant rise in the optical constants can be observed.

Additionally, the effective optical constants were numerically simulated using the Maxwell-Garnett (MG) approach, where spherical Ir NPs are embedded in a host (air) matrix. Figure 2b provides a rough estimation of these dispersion profiles showing an increasing trend in the refractive index n as the Ir surface coverage increases (see also SI). However, the extinction profiles were not suitably predicted by the MG model. Furthermore, a Bruggeman (BG) framework was applied to determine the effective optical constant of the samples, where the BG model assumes the Ir NP system as a homogeneous mixture of Ir and air. The BG simulations result in a better agreement with the experimental data. Especially for Ir coatings above the percolation threshold, as the MG approach is limited for a higher fill fraction of Ir inclusion. In addition, the large extinction coefficient of metallic inclusions and the resulting strong contrast in the refractive index to the host material lead to discrepancies in EMA theories. A detailed comparison of the dispersion profiles determined by SE measurements, MG and BG simulations is included in the SI. The growth of iridium in the nucleation regime depends very much on the process conditions, and slight deviations in the size of the particles result in significant variations in the reflectance, transmittance, and optical losses.

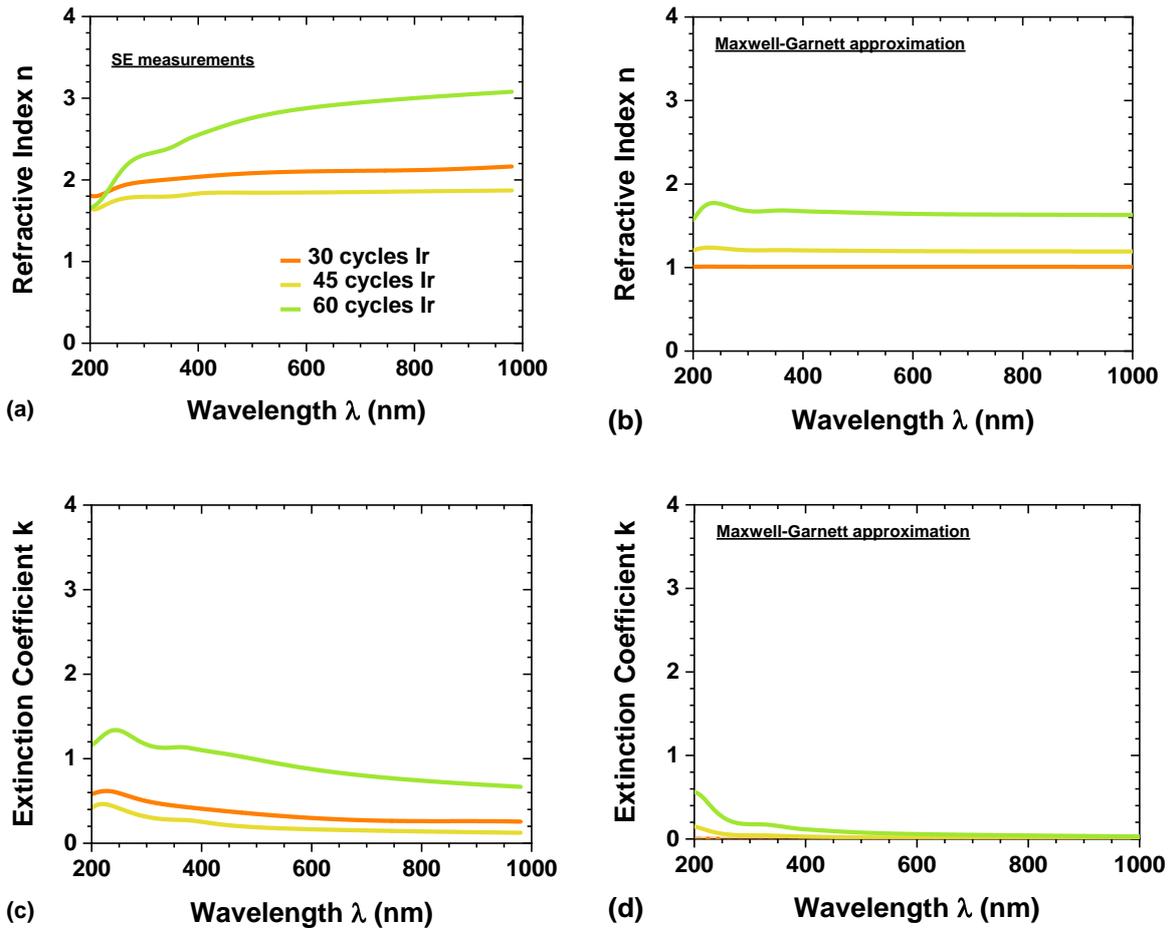

**Figure 2**. Optical constants of Ir coatings with different numbers of ALD cycles from 200 to 1000 nm wavelength. The refractive index n and extinction coefficient k were evaluated by fitting (a, c) spectral ellipsometry (SE) measurements and modeled using (b, d) Maxwell-Garnett (MG) theory.

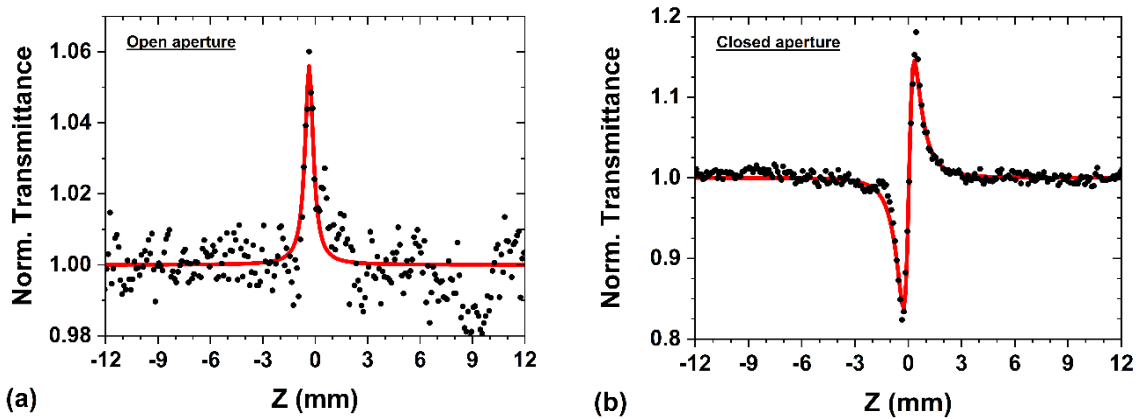

**Figure 3**. Normalized transmittance of Z scan measurements for (a) open aperture and (b) closed aperture for an Ir coating with 45 cycles.

The third-order nonlinearity of Ir NP layers of 30, 45, and 60 ALD cycles was then investigated experimentally using the Z-scan technique. Figure 3 shows a typical measurement for an Ir NP coating with 45 cycles. The normalized open aperture (OA) measurement shows a weak positive peak at around Z = 0, while no OA signal can be found from bare FS. The positive OA signal corresponds to saturable absorption or two-photon absorptions. No clear OA signal can be found for lower laser intensity. This result can be fitted as follows:

$$T(x) = \sum_{m=0}^{\infty} \frac{[-q_0]^m}{(m+1)^{3/2}} \quad , \quad q_0 = \frac{\beta I_0 L_{eff}}{1+x^2} \qquad 10,11$$

where $x=z/z_0$, $z_0$ is the diffraction length of a Gaussian laser beam, $\beta$ the nonlinear absorption coefficient, $L_{eff} = (1 - e^{-\frac{\alpha_0}{L}})/\alpha_0$ the sample's effective length, $\alpha_0$ the sample absorption coefficient, and L is the sample thickness, both of which can be extracted from the linear measurement results. The fitted $\beta$ value for 45 cycles sample is -3.35*10$^{-8}$ m/W.

In the CA measurement, the derivative shape curve originates from the nonlinear Kerr effect of the material. As shown in Fig. 3b., the normalized CA transmittance first decreases for the sample at the position before the focus (Z<0) and then increases after the focus (Z>0). This curve indicates a positive Kerr effect, corresponding to self-focusing in the sample. Normalized CA transmittance can be fitted with the following formula [47]:

$$T(x) = 1 + \frac{4x\Delta\Phi}{(1+x^2)(9+x^2)} + \frac{4(3x^2-5)\Delta\Phi^2}{(1+x^2)^2(9+x^2)(25+x^2)} + \frac{32(3x^2-11)x\Delta\Phi^3}{(1+x^2)^3(9+x^2)(25+x^2)(49+x^2)} \qquad 12$$

where $\Delta\Phi$ is the on-axis phase shift and $\Delta\Phi = 2\pi/(\lambda n_2 I_0 L_{eff})$, $\lambda$ is the fundamental wavelength, $I_0$ is the on-axis intensity at the focus, and $n_2$ is the fitting parameter for the Kerr coefficient. Since the bare FS substrate contributes significantly to the CA signal, the actual $n_2$ value of the sample is obtained by subtracting the fitted $n_2$ value of the substrate measured separately as a reference (see SI). The $n_2$ of the sample can be obtained as 1.86*10$^{-15}$ m$^2$/W, and the one measured from the bare FS substrate is around ~-2.11*10$^{-20}$ m$^2$/W. The measured $n_2$ value of the bare FS agrees well with the results reported [48,49], which could verify the fidelity of our measurement. The relatively large $n_2$ arises from the Ir NP sample. The thermal effect could affect the measured $n_2$ value; note that our laser pulse duration is about 100 fs, where the thermal effect arising from the acoustic wave could be neglected [50].

The real part and imaginary part of the third-order susceptibility $\chi^{(3)}$ of the samples can be calculated in SI units:

$$\chi_{Re}^{(3)} = \left(\frac{4}{3}\right) n_0 \varepsilon_0 c (n_0 n_2 - k_0 k_2) \qquad 13$$

$$\chi_{Im}^{(3)} = \left(\frac{4}{3}\right) n_0 \varepsilon_0 c (n_0 k_2 + k_0 n_2) \qquad 14$$

where $n_0$ and $k_0 = \lambda\alpha/4\pi$ are the linear refractive index and extinction coefficient, $k_2 = \lambda\beta/4\pi$ is the nonlinear extinction coefficient. The experimentally determined $n_2$, $\beta$, and $\chi^{(3)}$ values of the measured three samples are given in Table 2.

**Table 2**. Experimental and simulated values of nonlinear refractive index $n_2$, nonlinear absorption coefficient $\beta$, and third order susceptibility $\chi^{(3)}$ of selected Ir NP samples at 700 nm fundamental wavelength.

| ID | Number of ALD Cycles | Nonlinear Refractive Index $n_2$ (10$^{-15}$ m$^2$/W) | $n_2$ (10$^{-15}$ m$^2$/W) | Nonlinear Absorption Coefficient $\beta$ (10$^{-8}$ m/W) | $\beta$ (10$^{-8}$ m/W) | Ir Susceptibility (10$^{-17}$ m$^2$/V$^2$) (Measurements) | | Ir Susceptibility (10$^{-17}$ m$^2$/V$^2$) (Simulations) | |
|---|---|---|---|---|---|---|---|---|---|
| | | | | | | Re($\chi^{(3)}$) | Im($\chi^{(3)}$) | Re($\chi^{(3)}$) | Im($\chi^{(3)}$) |
| 1 | 30 | 1.77 ± 0.11 | 4.22 | -4.01 ± 0.33 | -7.44 | 3.24 ± 0.21 | -3.17 ± 0.27 | 3.52 | -3.41 |
| 2 | 45 | 1.86 ± 0.10 | 2.15 | -3.35 ± 0.27 | -3.52 | 2.44 ± 0.14 | -2.08± 0.17 | 2.66 | -2.24 |
| 3 | 60 | 1.69 ± 0.13 | 2.95 | -2.88 ± 0.12 | -3.54 | 6.53 ± 0.35 | -3.53± 0.32 | 7.28 | -3.80 |

Simulations of the third-order nonlinearity of Ir NP layers were performed in the framework of nonlinear MG theory at wavelengths following the Z-scan measurements. Both the bulk nonlinearity in terms of the third order susceptibility eq. (4) and the effective susceptibility from MG theory eq. (5) do not show a substantial deviation with particle size and can safely be assumed to be constant at the fundamental wave (FH, 700 nm) far from resonance and the third harmonic (TH, 233.3 nm), see Figs. 4(a, b) for the MG result and Fig. 4(c) for the bulk susceptibility of the three considered sample sizes. The NP size distribution in the films should be accounted for when particle sizes are above 10 nm or particularly close to the local surface plasmon resonance. Due to the amorphous mixing of Ir particles with air on a FS substrate, the effective third-order nonlinear susceptibility is strongly reduced in films with low particle density, represented by the volume fill fraction *f*. Here, we consider the three films with the lowest film thickness according to Table 1. Increasing the fill fraction further leads to unphysical results from this EMA, as already seen for the linear results. Fully closed films are correctly represented by the $f = 1$ (bulk) limit. Larger fill fractions, as shown in Fig. 4(b), yield a stronger nonlinear response (real part of the susceptibility, solid lines); however, the absorption (imaginary part of the susceptibility, dashed lines) is also increased and added to the linear permittivity with the third power of the local fields. At some larger NP density, the absorption starts to dominate. Our comparative study reveals that the theoretical calculations predict a $15 \times 10^5$-fold increase of the bulk susceptibility at the fundamental wavelength (700 nm) if amorphous gold films were used instead of iridium. In contrast, an enhancement for the iridium films grown with 60 ALD cycles is predicted around 233 nm wavelength. Compared to the fundamental wavelength applied in these experiments, the effective susceptibility using the current TH wavelength ($\approx$ 233 nm) is about $5 \times 10^6$ fold larger than at 700 nm. These differences between the metals arise in the local field enhancement at the FH. While gold reaches enhancement factors ~100, the enhancement factors of iridium in eq. 5 are of the order of ~1…35 at 700 nm wavelength. More cases are discussed in the supporting material and in Ref. [32].

We explore the influence of a thin coating around the Ir NPs by impurities such as a water film (n=1.33) or a shell of CH2-CH-CHO (acrylic aldehyde, n=1.367) in Fig. 4(d). The upper panel shows the dependence of the local surface plasmon resonance of Ir NPs in air with varying sizes. The uncoated Ir NP (black curves) shows negligible dependence on the size as discussed above. This is the classical Mie result within the local response approximation (LRA). Quantum size and confinement effects can be included through nonlocal optical response (red curves, NOR) [51]. This leads to a blue shift in the resonance position (red curves) and effectively reduces the third-order susceptibility. However, the effect of a thin 0.5 nm coating on the resonance position is much stronger than the finite size effect. The refractive indices of pure water and the considered CHO compound are close to each other, so the red-shift observed from the Ir-core-shell structures is similar (blue and green curves). Through eq. 4, this resonance shift enters the third-order nonlinear susceptibility, which is shown for the low-frequency limit following Miller's rule in the lower panel of Fig. 4(d). For the smallest Ir NP size, the presence of an aqueous coating could increase the susceptibility by up to one order of magnitude, pushing the Mie resonance of Ir NPs closer to the wavelength of the incoming laser field. Thus, the considered coating brings the susceptibilities closer to the experimentally observed values than the standard approach and hints towards the importance of considering a thin film of water forming on the Ir NPs exposed in air.

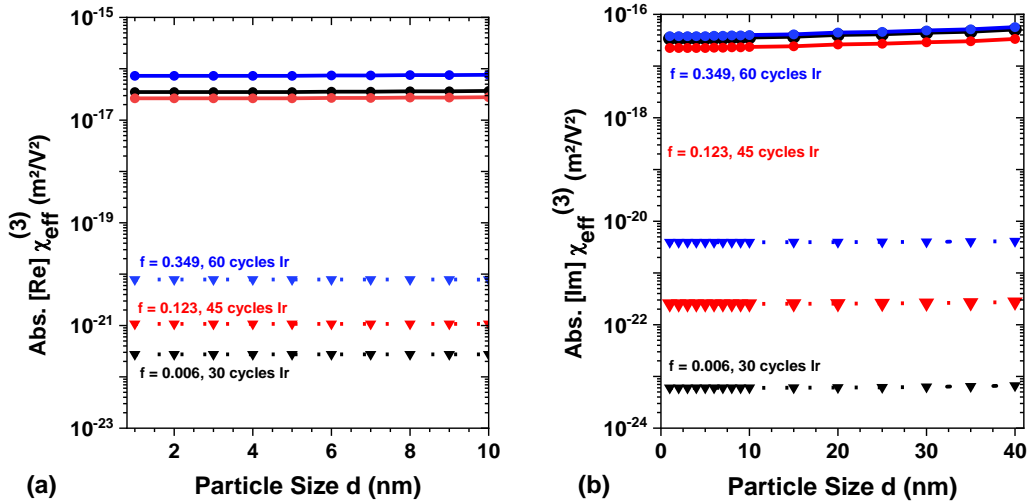

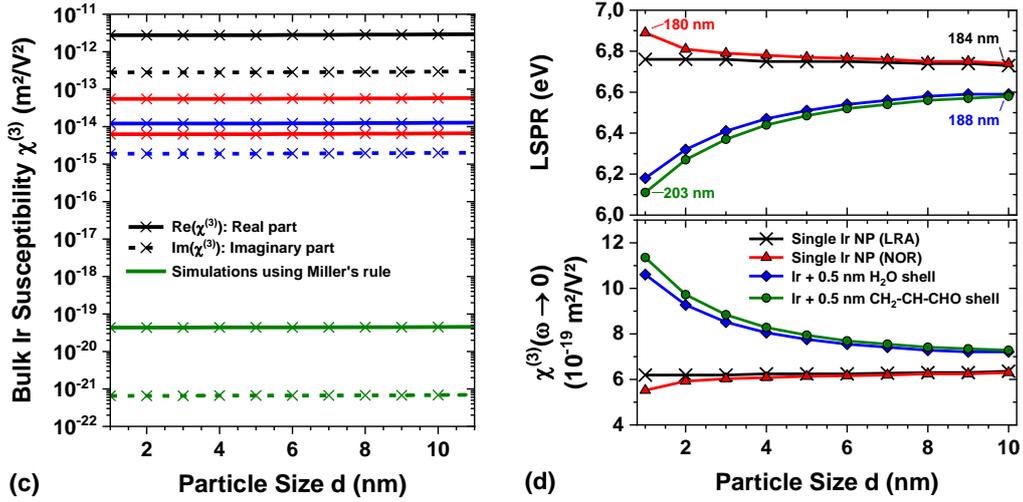

**Figure 4**. (a) Real and (b) imaginary part of the effective susceptibility at 700 nm FH for Ir coatings depending on the particle size. The solid curves represent the values with experimentally determined $\chi^{(3)}$ and fill fraction values, and the dashed curve is derived from the Miller's rule. (c) Bulk Ir susceptibility, the real (solid) and imaginary (dashed) parts are shown for data fitted to the three experimental results, yielding three different nonlinear coefficients $a=b\omega_p^8$, with $b = (-1.12 – i0.15) \times 10^{-15}$ m²/V² blue curves), $b = (-5.08 – i0.46) \times 10^{-15}$ m²/V² (red curves), $b = (-2.55 – i0.2) \times 10^{-13}$ m²/V² (black curves), and the result according to Miller's rule (green curves). (d) Shift of the localized surface plasmon resonance (LSPR, top) and Ir susceptibility according to Miller's rule (bottom), affected by an ultra-thin shell of water (blue) and $CH_2$-CH-CHO (green) surrounding the Ir nanoparticles (NP) in air. In addition, we compared single Ir NP in the classical local response approximation (LRA, black) and the size-induced quantum shift from the nonlocal optical response (NOR, red).

Two central problems arise in this theoretical description. Firstly, this approach to nonlinear optical properties of amorphous Ir thin films is instantaneous, and the shape and duration of the excitation pulse are not accounted for. Studies on gold films [52] show that the length of the pulse duration has a tremendous impact on the third order nonlinear susceptibility shifting it over orders of magnitude for longer pulses. However, our theoretical result, being instantaneous, is at the limit of vanishing pulse duration. Several orders of magnitude differences between theoretical results and measurements are, thus, not surprising. This could be amended by fitting the power law observed in gold [52] to the case of the discussed Ir films by measuring their nonlinear optical response for varying pulse durations and accounting for this effect in calculating the nonlinear susceptibility.

Second, the 233.3 nm wavelength is very close to the Mie resonance of nano-sized Ir particles. Thus, the effective Ir susceptibility values are largely increased compared to the off-resonant case at the fundamental wavelength of 700 nm and can become even larger than the corresponding bulk values. Additionally, the amorphous mixing in the Maxwell-Garnett theory at very low fill factors is dominated by the nonlinearity of the host, which defines a lower limit. Air has a nonlinear coefficient several orders of magnitude lower than the FS substrate; however, the influence of a substrate cannot be accounted for within the MG theory. As the studied Ir NP layers are a few nanometers thick, their interaction with the substrate cannot be neglected and results in Figs. 4(a)-(c) are for Ir nanoparticles at the FS/air interface. A self-consistent approach to the nonlinear optical properties of thin films of amorphous composites depending on their thickness was recently realized in our work [32].

Extensive studies have been conducted on different Au nanoscale systems by the Z-scan technique indicating, e.g., $n_2$ of Au NP in colloid [53] shows ≈ 1 to 2x $10^{-17}$ m²/W, Au-$SiO_2$ composite system [54,55] indicates $n_2$ of 0.4 to 8.9x $10^{-14}$ m²/W, periodic nanostructured film [56] of 14 nm gives rise to $n_2$ of 9.2x$10^{-13}$ m²/W. An increase in $n_2$ has been observed in Au NP aggregate film (≈ 9.2x$10^{-13}$ m²/W) due to local field enhancement as compared to 8 nm Au film [57] possessing $n_2$ of 6.5x$10^{-13}$ m²/W. Various studies reported on the nonlinear optical response from a wide range of nanoscale materials, for instance, large $n_2$ (6x $10^{-14}$ m²/W) observed in an ITO layer at epsilon-near-zero conditions [16], 2D $WS_2$ [47] shows $n_2$ of about 8x $10^{-13}$ m²/W, the two-photon absorption coefficient of $TiO_2$ thin films [58] is about 2-5x$10^{-13}$ m/W, 5 nm Au thin film [59] possesses nonlinear absorption coefficient as large as of ≈ 1.9 to 5.3x$10^{-5}$ m/W, 2D BiOBr nanoflakes display $n_2$ ≈ $10^{-14}$ m²/W and ß ≈$10^{-7}$ m²/W, respectively [60]. Hence, compared with other nanoscale materials, the third order nonlinear optical response of 3-5 nm Ir coatings is of decent strength. With the incorporation of such ultrathin Ir NP in various dielectric matrices one could pave the way to signal enhancement. Currently, such composite materials are being investigated for both second and third harmonic generation at VIS/NIR wavelengths.

While the nonlinear refractive index of Ir nanoparticles at 700 nm wavelength ($n_2$ ~$10^{-15}$ m$^2$/W) is bit lower than for the above-mentioned materials, a promising enhancement is predicted when using a deep UV (DUV) fundamental wavelength at 233 nm. Noteworthy, such large nonlinear optical constants are obtained for the above-mentioned materials at particular wavelengths with a strong enhancement condition. Epsilon near zero (ENZ) conditions are typically found to lead to these large nonlinearities, while off-resonance the values are 3-4 order of magnitude lower. The Mie resonance condition of Ir possibly enhances the nonlinear optical properties at DUV wavelengths. However, the expected TH wavelength is in the vacuum UV range and such measurements cannot be carried out currently in our laboratories. Further development in high power, ultrashort pulsed UV lasers would open new possibilities to investigate such novel material systems. The search for new, and stable material systems is essential to address these applications.

# 4  Conclusions

In this work, we systematically investigated the linear and nonlinear optical properties of Ir NP coatings grown by atomic layer deposition. Various spectroscopic and microscopic techniques have been applied to probe the linear optical properties of Ir NP systems. The variation in linear optical properties (e.g., optical spectra, dispersion profiles) is evident based on the nanoparticle density of Ir NP coatings. Further, we implemented a combined experimental and theoretical approach to estimate the third-order susceptibility of Ir NP coatings. A decent third-order nonlinearity (≈ $10^{-17}$ m$^2$/V$^2$) was observed using the Z-scan technique with 3 - 5 nm Ir functionalization. This work demonstrates the potential of ultrathin Ir NPs as an alternative nonlinear optical material at an atomic scale. This opens a gateway to develop new engineered materials, for instance, based on incorporating Ir within dielectric matrices and exploring the tunability of material properties. Increasing interest in DUV and VUV nonlinear applications might prompt the use of such materials.

**Corresponding Author**
*adriana.szeghalmi@iof.fraunhofer.de

**Funding Sources**
We acknowledge support by the Fraunhofer Society Attract Project (066-601020), Fraunhofer IOF Center of Excellence in Photonics, and the Deutsche Forschungsgemeinschaft (DFG, German Research Foundation) Collaborative Research Center (CRC/SFB) 1375 "NOA – Nonlinear Optics down to Atomic scales" (398816777). Paul Schmitt thanks the Thüringer Aufbaubank (TAB) for promoting his doctoral research studies. Weiwei Li and Matthias F. Kling acknowledge support by the Max Planck Society via the IMPRS for Advanced Photon Science, and the Max Planck Fellow program, respectively. Zilong Wang is grateful for support by the Alexander von Humboldt foundation.

**ACKNOWLEDGMENT**
The authors thank David Kästner for his technical support.

**REFERENCES**

1. M. Taghinejad and W. Cai, "All-Optical Control of Light in Micro- and Nanophotonics," ACS Photonics **6**, 1082–1093 (2019).
2. J. Kurumida and S. J. B. Yoo, "Nonlinear Optical Signal Processing in Optical Packet Switching Systems," IEEE Journal of Selected Topics in Quantum Electronics **18**, 978–987 (2012).
3. A. S. Dvornikov, E. P. Walker, and P. M. Rentzepis, "Two-Photon Three-Dimensional Optical Storage Memory," The Journal of Physical Chemistry A **113**, 13633–13644 (2009).
4. F. Walter, G. Li, C. Meier, S. Zhang, and T. Zentgraf, "Ultrathin Nonlinear Metasurface for Optical Image Encoding," Nano Letters **17**, 3171–3175 (2017).

5. S. Palomba and L. Novotny, "Near-Field Imaging with a Localized Nonlinear Light Source," Nano Letters **9**, 3801–3804 (2009).
6. R. Gadhwal and A. Devi, "A review on the development of optical limiters from homogeneous to reflective 1-D photonic crystal structures," Optics & Laser Technology **141**, 107144 (2021).
7. X. Zhang, D. Zhang, D. Tan, Y. Xian, X. Liu, and J. Qiu, "Highly Defective Nanocrystals as Ultrafast Optical Switches: Nonequilibrium Synthesis and Efficient Nonlinear Optical Response," Chem. Mater. **32**, 10025–10034 (2020).
8. G. Liu, S. You, M. Ma, H. Huang, and N. Ren, "Removal of Nitrate by Photocatalytic Denitrification Using Nonlinear Optical Material," Environmental Science & Technology **50**, 11218–11225 (2016).
9. B. Kulyk, Z. Essaidi, V. Kapustianyk, B. Turko, V. Rudyk, M. Partyka, M. Addou, and B. Sahraoui, "Second and third order nonlinear optical properties of nanostructured ZnO thin films deposited on a-BBO and LiNbO3," Optics Communications **281**, 6107–6111 (2008).
10. Y.-x. Zhang and Y.-h. Wang, "Nonlinear optical properties of metal nanoparticles: a review," RSC Adv **7**, 45129–45144 (2017).
11. J. W. You, S. R. Bongu, Q. Bao, and N. C. Panoiu, "Nonlinear optical properties and applications of 2D materials: theoretical and experimental aspects," Nanophotonics **8**, 63–97 (2019).
12. D. Lehr, J. Reinhold, I. Thiele, H. Hartung, K. Dietrich, C. Menzel, T. Pertsch, E.-B. Kley, and A. Tünnermann, "Enhancing second harmonic generation in gold nanoring resonators filled with lithium niobate," Nano Letters **15**, 1025–1030 (2015).
13. O. Sánchez-Dena, P. Mota-Santiago, L. Tamayo-Rivera, E. V. García-Ramírez, A. Crespo-Sosa, A. Oliver, and J.-A. Reyes-Esqueda, "Size-and shape-dependent nonlinear optical response of Au nanoparticles embedded in sapphire," Opt. Mater. Express **4**, 92 (2014).
14. R. Sato, M. Ohnuma, K. Oyoshi, and Y. Takeda, "Experimental investigation of nonlinear optical properties of Ag nanoparticles: Effects of size quantization," Phys. Rev. B **90** (2014).
15. F. Che, S. Grabtchak, W. M. Whelan, S. A. Ponomarenko, and M. Cada, "Relative SHG measurements of metal thin films: Gold, silver, aluminum, cobalt, chromium, germanium, nickel, antimony, titanium, titanium nitride, tungsten, zinc, silicon and indium tin oxide," Results in Physics **7**, 593–595 (2017).
16. M. Z. Alam, I. de Leon, and R. W. Boyd, "Large optical nonlinearity of indium tin oxide in its epsilon-near-zero region," Science (New York, N.Y.) **352**, 795–797 (2016).
17. A. Wickberg, C. Kieninger, C. Sürgers, S. Schlabach, X. Mu, C. Koos, and M. Wegener, "Second-Harmonic Generation from ZnO/Al 2 O 3 Nanolaminate Optical Metamaterials Grown by Atomic-Layer Deposition," Advanced Optical Materials **4**, 1203–1208 (2016).
18. L. Alloatti, C. Kieninger, A. Froelich, M. Lauermann, T. Frenzel, K. Köhnle, W. Freude, J. Leuthold, M. Wegener, and C. Koos, "Second-order nonlinear optical metamaterials: ABC-type nanolaminates," Appl. Phys. Lett. **107**, 121903 (2015).
19. A.-C. Probst, M. Stollenwerk, F. Emmerich, A. Büttner, S. Zeising, J. Stadtmüller, F. Riethmüller, V. Stehlíková, M. Wen, L. Proserpio, C. Damm, B. Rellinghaus, and T. Döhring, "Influence of sputtering pressure on the nanostructure and the X-ray reflectivity of iridium coatings," Surf. Coat. Technol. **343**, 101–107 (2018).
20. P. L. Henriksen, D. D. M. Ferreira, S. Massahi, M. C. Civitani, S. Basso, J. Vogel, J. R. Armendariz, E. B. Knudsen, I. G. Irastorza, and F. E. Christensen, "Iridium thin-film coatings for the BabyIAXO hybrid X-ray optic," Applied Optics **60**, 6671–6681 (2021).

21. J. W. Arblaster, "Crystallographic Properties of Iridium," Platinum Met. Rev. **54**, 93–102 (2010).
22. J. Vila-Comamala, S. Gorelick, E. Färm, C. M. Kewish, A. Diaz, R. Barrett, V. A. Guzenko, M. Ritala, and C. David, "Ultra-high resolution zone-doubled diffractive X-ray optics for the multi-keV regime," Opt. Express **19**, 175–184 (2011).
23. T. Weber, T. Käsebier, A. Szeghalmi, M. Knez, E.-B. Kley, and A. Tünnermann, "Iridium wire grid polarizer fabricated using atomic layer deposition," Nanoscale Res Lett **6**, 558 (2011).
24. R. Hemphill, M. Hurwitz, and M. G. Pelizzo, "Osmium atomic-oxygen protection by an iridium overcoat for increased extreme-ultraviolet grating efficiency," Applied Optics **42**, 5149–5157 (2003).
25. P. Schmitt, N. Felde, T. Döhring, M. Stollenwerk, I. Uschmann, K. Hanemann, M. Siegler, G. Klemm, N. Gratzke, A. Tünnermann, S. Schwinde, S. Schröder, and A. Szeghalmi, "Optical, structural, and functional properties of highly reflective and stable iridium mirror coatings for infrared applications," Opt. Mater. Express (2021).
26. A. Colombo, C. Dragonetti, V. Guerchais, C. Hierlinger, E. Zysman-Colman, and D. Roberto, "A trip in the nonlinear optical properties of iridium complexes," Coordination Chemistry Reviews **414**, 213293 (2020).
27. L. Yan and Woollam, John, A., "Optical constants and roughness study of dc magnetron sputtered iridium films," J. Appl. Phys. **92**, 4386–4392 (2002).
28. S. Kohli, D. Niles, C. D. Rithner, and P. K. Dorhout, "Structural and optical properties of Iridium films annealed in air," Adv. X-Ray Anal. **45**, 352–358 (2002).
29. L. Ghazaryan, K. Pfeiffer, P. Schmitt, V. Beladiya, S. Kund, and A. Szeghalmi, "Atomic Layer Deposition," in *digital Encyclopedia of Applied Physics* (Wiley‐VCH Verlag, 2020), pp. 1‐44.
30. P. Paul, M. G. Hafiz, P. Schmitt, C. Patzig, F. Otto, T. Fritz, A. Tünnermann, and A. Szeghalmi, "Optical bandgap control in Al2O3/TiO2 heterostructures by plasma enhanced atomic layer deposition: Toward quantizing structures and tailored binary oxides," Spectrochimica acta. Part A, Molecular and biomolecular spectroscopy **252**, 119508 (2021).
31. P. Schmitt, V. Beladiya, N. Felde, P. Paul, F. Otto, T. Fritz, A. Tünnermann, and A. V. Szeghalmi, "Influence of Substrate Materials on Nucleation and Properties of Iridium Thin Films Grown by ALD," Coatings **11**, 173 (2021).
32. N. Daryakar and C. David, "Thin Films of Nonlinear Metallic Amorphous Composites," Nanomaterials **12**, 3359 (2022).
33. P. Genevée, E. Ahiavi, N. Janunts, T. Pertsch, M. Oliva, E.-B. Kley, and A. Szeghalmi, "Blistering during the atomic layer deposition of iridium," J. Vac. Sci. Technol. A **34**, 01A113 (2016).
34. "ImageJ," https://imagej.nih.gov/ij/.
35. S. Schröder, T. Herffurth, H. Blaschke, and A. Duparré, "Angle-resolved scattering: an effective method for characterizing thin-film coatings," Applied Optics **50**, C164‐C171 (2011).
36. V. A. Markel, "Introduction to the Maxwell Garnett approximation: tutorial," JOSA A **33**, 1244–1256 (2016).
37. Y. Battie, A. Resano-Garcia, N. Chaoui, Y. Zhang, and A. En Naciri, "Extended Maxwell-Garnett-Mie formulation applied to size dispersion of metallic nanoparticles embedded in host liquid matrix," The Journal of Chemical Physics **140**, 44705 (2014).
38. J. E. Sipe and R. W. Boyd, "Nonlinear susceptibility of composite optical materials in the Maxwell Garnett model," Physical Review A **46**, 1614–1629 (1992).

39. K.-E. Peiponen, M. O. A. Mäkinen, J. J. Saarinen, and T. Asakura, "Dispersion Theory of Liquids Containing Optically Linear and Nonlinear Maxwell Garnett Nanoparticles," Optical Review **8**, 9–17 (2001).
40. J. J. Saarinen, E. M. Vartiainen, and K.-E. Peiponen, "On Tailoring of Nonlinear Spectral Properties of Nanocomposites Having Maxwell Garnett or Bruggeman Structure," Optical Review **10**, 111–115 (2003).
41. R. W. Boyd, R. J. Gehr, G. L. Fischer, and J. E. Sipe, "Nonlinear optical properties of nanocomposite materials," Pure and Applied Optics: Journal of the European Optical Society Part A **5**, 505–512 (1996).
42. A. Marini, M. Conforti, G. Della Valle, H. W. Lee, T. X. Tran, W. Chang, M. A. Schmidt, S. Longhi, P. S. J. Russell, and F. Biancalana, "Ultrafast nonlinear dynamics of surface plasmon polaritons in gold nanowires due to the intrinsic nonlinearity of metals," New Journal of Physics **15**, 13033 (2013).
43. R. W. Boyd, *Nonlinear Optics*, 2nd Edition (Elsevier, 2003).
44. R. del Coso and J. Solis, "Relation between nonlinear refractive index and third-order susceptibility in absorbing media," J. Opt. Soc. Am. B **21**, 640 (2004).
45. R. A. Maniyara, D. Rodrigo, R. Yu, J. Canet-Ferrer, D. S. Ghosh, R. Yongsunthon, D. E. Baker, A. Rezikyan, García de Abajo, F. Javier, and V. Pruneri, "Tunable plasmons in ultrathin metal films," Nat. Photonics **13**, 328–333 (2019).
46. N. Formica, D. S. Ghosh, A. Carrilero, T. L. Chen, R. E. Simpson, and V. Pruneri, "Ultrastable and atomically smooth ultrathin silver films grown on a copper seed layer," ACS applied materials & interfaces **5**, 3048–3053 (2013).
47. X. Zheng, Y. Zhang, R. Chen, X.'a. Cheng, Z. Xu, and T. Jiang, "Z-scan measurement of the nonlinear refractive index of monolayer WS(2)," Opt. Express **23**, 15616–15623 (2015).
48. D. Milam, "Review and assessment of measured values of the nonlinear refractive-index coefficient of fused silica," Applied Optics **37**, 546–550 (1998).
49. C. B. Schaffer, A. Brodeur, and E. Mazur, "Laser-induced breakdown and damage in bulk transparent materials induced by tightly focused femtosecond laser pulses," Meas. Sci. Technol. **12**, 1784–1794 (2001).
50. Y. Guillet, M. Rashidi-Huyeh, and B. Palpant, "Influence of laser pulse characteristics on the hot electron contribution to the third-order nonlinear optical response of gold nanoparticles," Phys. Rev. B **79** (2009).
51. A. Moradi, "Maxwell-Garnett effective medium theory: Quantum nonlocal effects," Physics of Plasmas **22**, 42105 (2015).
52. R. W. Boyd, Z. Shi, and I. de Leon, "The third-order nonlinear optical susceptibility of gold," Optics Communications **326**, 74–79 (2014).
53. H. P. S. Castro, H. Wender, M. A. R. C. Alencar, S. R. Teixeira, J. Dupont, and J. M. Hickmann, "Third-order nonlinear optical response of colloidal gold nanoparticles prepared by sputtering deposition," Journal of Applied Physics **114**, 183104 (2013).
54. F. Hache, D. Ricard, C. Flytzanis, and U. Kreibig, "The optical kerr effect in small metal particles and metal colloids: The case of gold," Appl. Phys. A **47**, 347–357 (1988).
55. A. L. Stepanov, "Nonlinear Optical Properties of Metal Nanoparticles in Silicate Glass," in *Glass nanocomposites. Synthesis, properties and applications / edited by Basudeb Karmakar, Klaus Rademann and Andrey Stepanov*, B. Karmakar, K. Rademann, and A. L. Stepanov, eds. (William Andrew, 2016), pp. 165–179.
56. H. Shen, B. Cheng, G. Lu, T. Ning, D. Guan, Y. Zhou, and Z. Chen, "Enhancement of optical nonlinearity in periodic gold nanoparticle arrays," Nanotechnology **17**, 4274–4277 (2006).

57. S. Bai, Q. Li, H. Zhang, X. Chen, S. Luo, H. Gong, Y. Yang, D. Zhao, and M. Qiu, "Large third-order nonlinear refractive index coefficient based on gold nanoparticle aggregate films," Appl. Phys. Lett. **107**, 141111 (2015).
58. O. Stenzel, S. Wilbrandt, C. Mühlig, and S. Schröder, "Linear and Nonlinear Absorption of Titanium Dioxide Films Produced by Plasma Ion-Assisted Electron Beam Evaporation: Modeling and Experiments," Coatings **10**, 59 (2020).
59. D. D. Smith, Y. Yoon, R. W. Boyd, J. K. Campbell, L. A. Baker, R. M. Crooks, and M. George, "z-scan measurement of the nonlinear absorption of a thin gold film," Journal of Applied Physics **86**, 6200–6205 (1999).
60. L. Jia, D. Cui, J. Wu, H. Feng, Y. Yang, T. Yang, Y. Qu, Y. Du, W. Hao, B. Jia, and D. J. Moss, "Highly nonlinear BiOBr nanoflakes for hybrid integrated photonics," APL Photonics **4**, 90802 (2019).